# Multimodal Optical Techniques in Pre-Clinical Evaluation of Oral Cancer: Fluorescence Imaging and Spectroscopic Devices

**Pramila Thapa[1], Veena Singh[1], Virendra Kumar[1], Sunil Bhatt[1], Kiran Maurya[2], Vivek Nayyar[2], Kiran Jot[2], Deepika Mishra[2], Anurag Shrivastava[3] and Dalip Singh Mehta[1*]**

[1]*Bio-photonics and Green-photonics Laboratory, Department of Physics, Indian Institute of Technology Delhi, Hauz-Khas, New Delhi 110016, India.*

[2]*Department of Oral Pathology and Microbiology, Center for Dental Education & Research, All India Institute of Medical Sciences (AIIMS), Ansari Nagar, New Delhi 110029, India.*

[2] *Department of Surgical Disciplines, All India Institute of Medical Sciences (AIIMS), Ansari Nagar, New Delhi 110029, India.*

**Correspondence \*** Dalip Singh Mehta, Bio-photonics and Green Photonics Laboratory, Department of Physics, Indian Institute of Technology Delhi, Hauz-Khas, New Delhi 110016, India. Email: mehtads@physics.iitd.ac.in

## Abstract

**Objective:** Survival rate of oral squamous cell carcinoma (OSCC) patients is very poor and can be improved using highly sensitive, specific and accurate techniques. Autofluorescence and fluorescence techniques are very sensitive and useful in cancer screening. Furthermore, fluorescence spectroscopy is directly linked with molecular levels of human tissue and can be used as quantitative tool for cancer detection.

**Materials and Methods:** Here, we report development of multi-modal autofluorescence and fluorescence imaging and spectroscopic (MAF-IS) smartphone-based systems for fast and real time oral cancer screening. Fluorescence-autofluorescence images and spectroscopic datasets shows significant change in oral cancer and normal tissue in terms of fluorescence-intensity, spectral-shape, and red-shift respectively.

**Results:** In this study, total 68 samples (33 cancerous and 35 normal) of 18 OSCC patients and 13 patients of precancerous tissues (dysplasia and fibrosis) are screened. Main remarkable finding of the study is presence of three peaks viz ~636 nm, ~680 nm and ~705 nm with decrease in intensity around 450 nm ~ 520 nm in OSCC in case of autofluorescence. Another finding is red shift in fluorescence spectroscopy of OSCC, dysplasia and fibrosis from normal which is 6.59±4.54 nm, 3±4.78 nm and 1.5±0.5 nm respectively and can be used as cancer marker in real-time screening. Finally, support vector machine (SVM) based classifier is applied for classification of OSCC tissue from normal tissue. The average sensitivity, specificity and accuracy are found as 88.89% ,100 %, and 95%, respectively.

**Conclusion:** Autofluorescence and fluorescence-based imaging and spectroscopy is used for pre-clinical screening of different oral lesions.

## 1. Introduction

Oral squamous cell carcinoma (OSCC) is the most frequent and common malignancy (around 84%-97%) among oral cancer[1]. Annually 77,000 people are affected by oral cancer and approximately 52,000 deaths are reported in India[2]. Oral cancer starts from altering the epithelial cells then it turns out to hyperplasia, thereafter dysplasia and ultimately if untreated, transforms into OSCC. The survival rate for the oral cancer patients in 5 years after various treatments comes around 50%[3]. However, survival rate can be increased with the help of early diagnosis and regular screening. Since conventional techniques such as histopathology are very lengthy, time-consuming, and tedious in nature[4]. Even for initial screening biopsy followed by histopathology required most of the time, which in itself is an invasive technique and amenable to subjective errors and requires experienced pathologists for diagnosis[4]. Therefore, there is an urgent need for techniques which are non-invasive, non –contact, fast and accurate and can-perform early diagnosis.

Numerous optical techniques which behave as non-invasive, fast, and accurate techniques have been developed and can be used for the screening of OSCC. Autofluorescence (AF), fluorescence (FL) imaging and spectroscopy, polarization imaging, optical coherence tomography (OCT), phase imaging etc. are used for the analysis of different properties of cancer tissues and cancer cell lines[5-11]. Polarization indicates change in intensity-based parameters such as degree of polarization, stokes parameters etc. for cancer and normal tissues which are not specific and depends upon illumination wavelength[12]. Phase imaging and OCT gives the phase information about cancerous and normal cells and tissues[7, 13], OCT also indicates the phase information with tomographic details of the tissues[14], though these techniques are quantitative but still lacks the information about different chemical compositions and molecular specificities within cancerous tissues[15]. AF and FL are molecular level techniques; usually, AF targets natural proteins or biomarkers present in human tissue and FL uses an exogenous agent. Among aforementioned optical techniques, AF and FL are most trending techniques for oral cancer screening, as in oral cavity normal tissues acts differently from malignant tissues in terms of proteomic biomarkers[16]. While all other techniques are bulky, need sample preparation, not specific and insufficient to tell about chemical compositions as well as formation of any proteins in cancerous environment which itself is a significant cancer biomarker[4]. AF targets naturally occurring endogenous fluorophores and proteins such as nicotinamide adenine di-nucleotide (NAD), Flavin di-nucleotide (FAD), collagen, porphyrins etc. which are main cancer biomarkers present in human body[16-18].

There are a number of devices which work on the principle of AF and are commercially available, such as velescope, oral scan etc.[19, 20]. These available devices are based on only AF imaging module and don't include spectroscopy making these devices less accurate in cancer screening[21]. AF spectroscopy plays an important role in oral cancer screening; as changes in epithelia composition, epithelia stromal and metabolism leads to the AF in oral cancer tissue[22]. Intracellular and extracellular changes in oral lesion leads to identify the different stages of progression with help of AF-FL spectroscopy[23]. AF and FL images offers qualitative information about disease progression and spectra offers quantitative molecular level changes based on endogenous and exogenous fluorophores. Endogenous and exogenous fluorophores are very specific (each molecule has separate electronic band and corresponding emission spectrum) in nature and can be used in real time cancer screening[24]. Methylene blue, toluidine blue, fluorescein are some exogenous fluorophores which are commonly used in cancer screening[8, 25]. Sensitivity and specificity of FL is very prominent and thus shows its capability in oral cancer screening[26].

All available FL and AF techniques for oral cancer screening uses imaging and spectroscopy individually[4]. In addition to single modal screening techniques, multi-modal techniques are also used for oral cancer screening. Multi-modality of a system enhances its performance and gives more accurate results rather than single modality. In this paper, novel multi-modal autofluorescence and fluorescence imaging and spectroscopic (MAF-IS) smartphone-based systems are developed for oral cancer screening. MAF-IS systems have four modalities i.e., AF imaging, spectroscopy, FL imaging, spectroscopy. AF offers native fluorescence from oral tissue giving a global image supported by a spectrum which is entirely different from normal tissue. Since AF signal is very low thus it is assisted with FL. FL provides good sensitivity by tagging the exogenous dyes with specific proteins and coenzymes giving global FL images and spectra [38] which makes systems more accurate. These techniques are non-contact, non-invasive and fast; integration of multimodality is unique and is demonstrated for the first time as according to authors knowledge. Conventionally all four methods are used independently and required systems are bulky, costly, needed sample preparation and skilled personal. In current healthcare scenario, there is a requirement to develop non-contact and non-invasive point-of-care screening devices which are small, accurate, portable, and inexpensive. Our MAF-IS based system accomplished all requirements can be attached with smartphone and operated by nurses for screening many patients quickly. As India is a hub of oral cancer and most of patients belongs to rural India; here the developed MAF-IS system comes in origin and gives a great advantage to such patients. The developed MAF-IS system is proficient to screen oral cancer patients at large scale with a high accuracy acquiring the multi-modality of the system.

## 2. Materials and Methods

### 2.1 Theoretical Background

FL and AF are well-known and well-established techniques for cancer screening. Here, two specific excitation wavelengths, 365 nm, and 405 nm are used for AF[27]. Motive of using two different excitations is acquiring the change in cancer progression along whole range of endogenous fluorophores present in human body. 365 nm is used for imaging which targets endogenous fluorophores NADH and FAD and collagen matrix[18]. 405 nm is used for spectroscopy, and probed NADH, FAD and porphyrins[16, 28]. Here in this manuscript, fluorescein sodium salt (FSS) is used as an exogenous agent for the cancer screening in FL as it is easily available and very cheap compared to others. FSS is dianion form of fluorescein (500 nm – 560 nm)[29]. One of the authors in[30] shows usage of FSS for first time and obtained 96.6 % sensitivity for OSCC detection. As fluorescence spectroscopy gives red shift for

OSCC compared to healthy tissue and high quantum yield in cancerous tissue, thus shows its potentiality in cancer screening[8]. Heterogeneous nature of cancer tissue is main reason for extra tagging of dye and obtained red-shift[8]. Image and spectra analysis for SVM can be found in supplementary file.

After extracting parameters from image and spectra analysis, support vector machine (SVM) is used for classification of OSCC tissue and normal tissue which provides an alternative for complex classification problems and gives great accuracy[31]. For current data interpretation, Machine Learning Toolbox of MATLAB R2020b is used. Here, 80% of data set is used randomly for training, and 20% of data is used for testing purpose. Receiver operating characteristic (ROC) curve is used for evaluating quality of SVM model.

## 2.2 Patients and Methods

A total 18 OSCC patients are included, 35 normal and 33 cancerous sample points were used. The patients were offered detailed printed information about study, and they signed "informed written consent". The study was approved by ethics committee of All India Institute of Medical Sciences (AIIMS) New Delhi. All measurements were recorded in-vivo procedure. Suspicious patients reported at general OPD and are referred for this study. The experimental procedure is shown in Fig. 1(a). The patients were subjected to clinical oral visual examination using 365nm LEDs for AF imaging and 405 nm laser for AF spectroscopy. This was followed by FL imaging and spectroscopy using FSS. FSS was topically applied with cotton tip applicator on the abnormal area and surrounding area. After 2 minutes, patient was told to do rinse mouth from water which removes excess dye and was examined under 488 nm and FL images and spectra are recorded. Suspicious area and normal area are demarked by clinician, and after that spectroscopy and imaging is done from the same area. The whole procedure is done in dark room. At the end, results are compared with histopathological results. Figure 1(b) indicates in-vivo and in-vitro procedure of patient's data collection with total time.

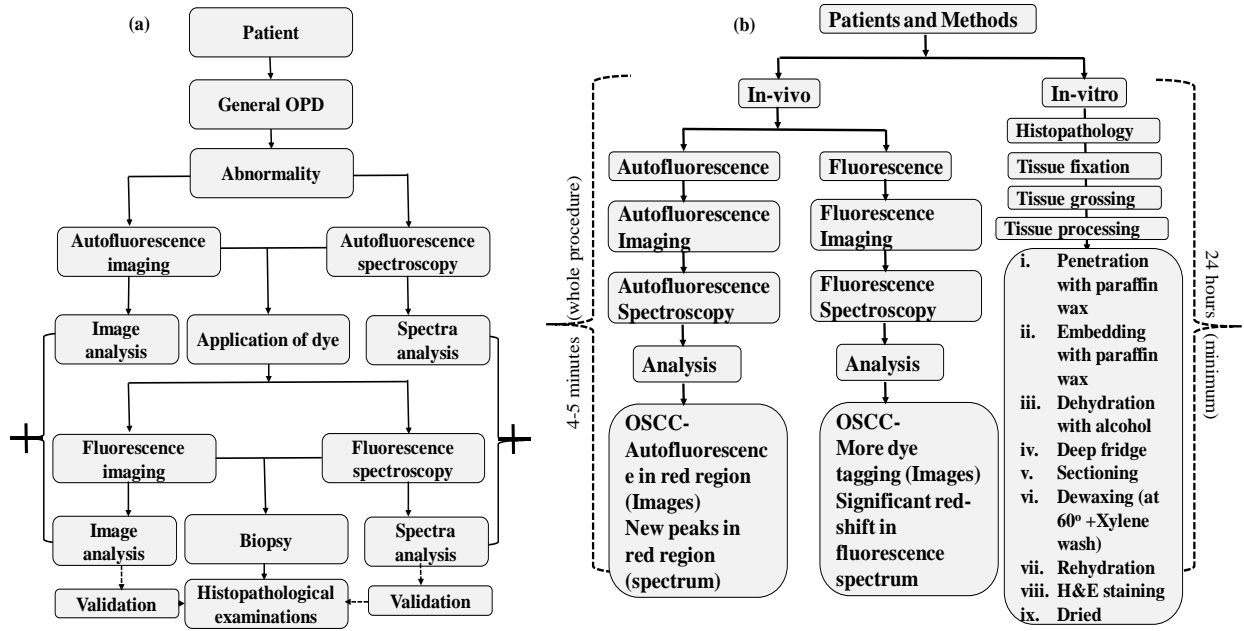

Figure 1. Flow charts of procedure. (a) Patient selection for autofluorescence and fluorescence study. (b) Whole procedure of patient's in-vivo and in-vitro sample collection at AIIMS hospital.

## 3. Experimental Setup

Developed MAF-IS systems are made up of two units, one is imaging device and second is spectroscopic device, which are assembled and developed in laboratory. Imaging devices are made up in form of gun-type for convenience of oral patients. AF and FL imaging devices has 10 LEDs (365 nm for AF and 488 nm for FL) on periphery of filter (long pass >420nm for AF, band pass filter 500-560 nm for FL) compartment. Webcam (Microsoft) is mounted at centre of device, then a filter is placed, which filters the AF and FL signal from captured images of patient. A mount is used for assembling LEDs, filter, webcam, polarizer, and analyzer. An intensity controller is attached with handle of device and an on/off button for LEDs is provided. Webcam can be accessed with help of smartphone, or we can use smartphone's camera for capturing the images of patients as according to clinician's preference and patient's oral condition (sometimes lesion is far away). Figure 2 (a) indicates the detailed information about developed device with attached optical components.

The webcam or smartphone captures AF/FL images from patients and intensity controller fixes intensity of source. Captured images are then filtered out with help of filters and stored on smartphone or computer. Figure 2 (b) indicates the ray diagram for imaging device, light from LEDs get diverged and then incident on oral tissue, after incidence, tissue get scattered, and webcam or smartphone will capture the images. L is length from the tissue to camera lenses and can be adjusted depending upon clinician's requirement. For present study this length is fixed for almost 10 cm. Figure. 2 (c) shows schematic of usage of device on patient.

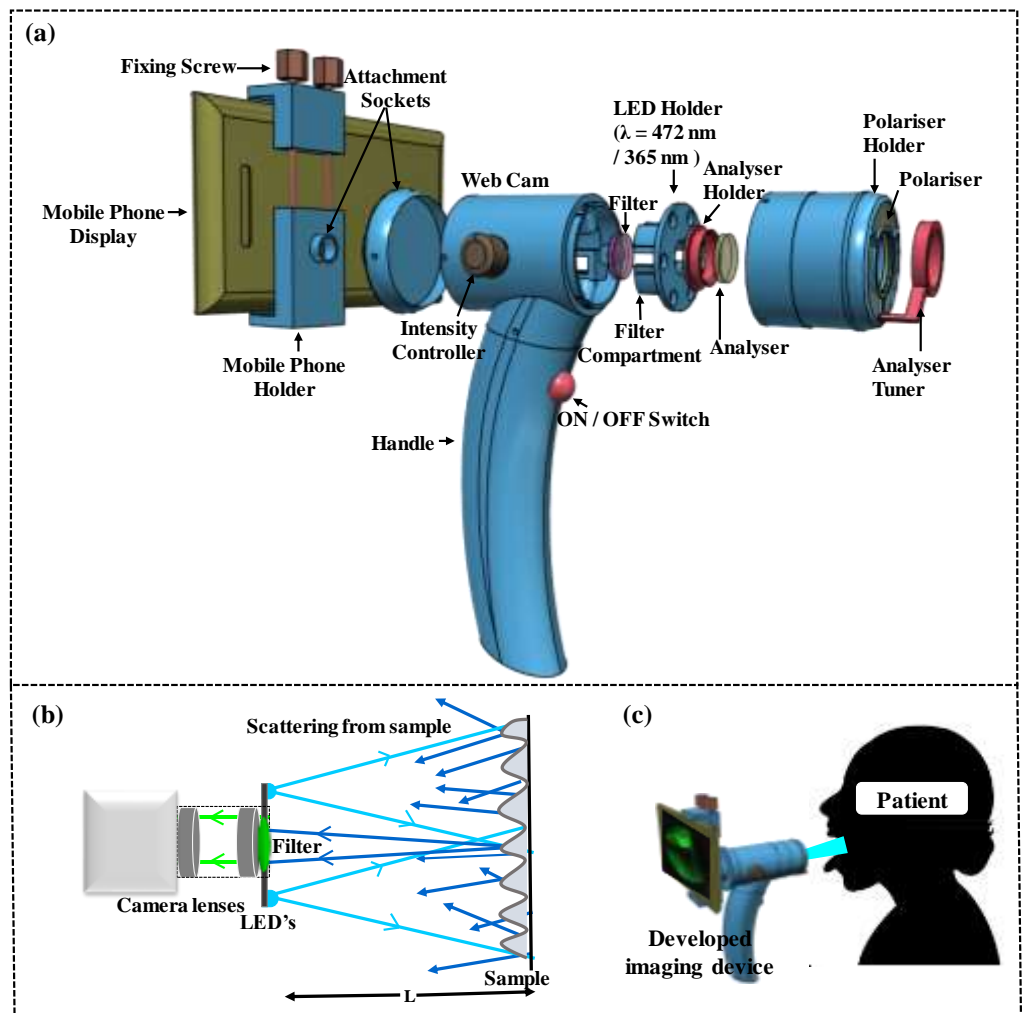

Figure 2. Imaging device. (a) Detailed information of developed imaging device. (b) Ray diagram for developed device. (c) Schematic of usage of developed device.

Spectroscopic device consists of two lasers; diode laser (488 nm, 3V DC, 5 mW, Sky Blue Dot Laser Diode Module Industrial Grade TY Laser) and a UV-A laser (405 nm, maximum power output < 5 mW), spectrometer (Lysen, 300 nm- 950nm) and a Y-coupler fibre (Avantes, FCR-7UVIR400-2-ME, 1503343) which is used for excitation and collection (total six fibers are used for excitation with 200 um core and collection fibre of 400 um). Fibre tip is inserted into disposable empty syringe which fixes length from tissue about 2 mm and then recorded AF and FL spectra. Figure 3 shows the experimental setup for spectroscopic device. Figure 3 (a) shows block diagram of spectroscopic components, Fig 3(b) shows geometry of fibre; central fibre is for collection and periphery fibres for excitation of the sample and Fig 3(c) shows the complete setup of fluorescence and autofluorescence spectroscopy. Laser power after fibre tip is ≤1 mW and patient's eyes were closed during procedure.

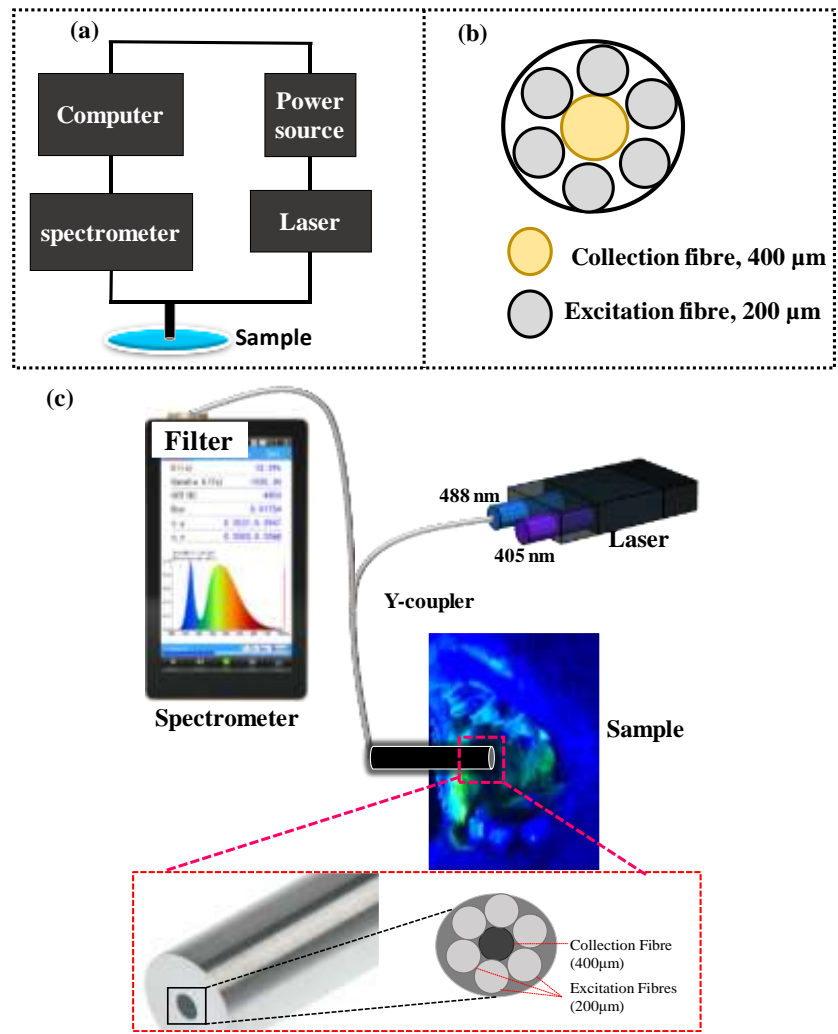

Figure 3. Spectroscopic device. (a) Block diagram for acquiring spectrum from the patients. (b) Schematic of y-coupler used. (c) Experimental setup for AF and FL spectroscopy of the oral cancer patients.

## 4. Results

Several parameters are extracted from AF and FL images and spectra. Figure 4 is showing AF and FL images and spectra of an OSCC patient. Figure 4 (a)-(c) are images indicating AF in red channel which corresponds protoporphyrin IX (PpIX), green and blue channels corresponding NADH-FAD crosslinks, respectively. Figure. 4(a) indicating good amount of AF of OSCC tissue in red channel (580~680 nm) which shows the presence of porphyrins, Fig.4 (b) indicates AF in green channel (480~600 nm) which shows the presence of FAD and in Fig4 (c) there is loss in AF in blue frame (400~500 nm) which shows loss in NADH-FAD crosslinks. Figure 4(g) is the AF spectrum of OSCC patient from 405 nm excitation and red line spectrum shows OSCC tissue's AF and green line spectrum shows normal tissue's AF in which spectral band 450~600 nm shows decrease in OSCC tissue compared to normal tissue which supports the AF getting from images. And there are three dominant peaks in OSCC tissue viz. 636 nm, 670 nm and 705 nm which are absent in normal tissue. These peaks correspond to excess amount of PpIX in OSCC. Detailed discussion of results is done in discussion part.

Figure 4 (d)-(f) are fluorescence images of the OSCC tissue where (d) indicates FSS tagging in OSCC tissue using 488nm, (e) indicates the digitally filtered fluorescent image which shows tagging of only altered tissue and (f) is intensity variation image of (e) indicating amount of fluorescence tagged within tissue. Figure 4 (e) clearly shows that OSCC is completely tagged with FSS and distinguish from surrounded normal area. Figure 4(f) shows within tissue there is some more tagging of dye which is due to the microenvironment of tumor as well as depends on individual patient. Further discussion of this is in discussion part. Figure 4(h) indicates the fluorescence spectra of cancerous tissue and normal tissue which shows that there is a 4 nm of red shift in cancerous tissue which supports the heterogeneity of cancerous tissue and has been explained in discussion part.

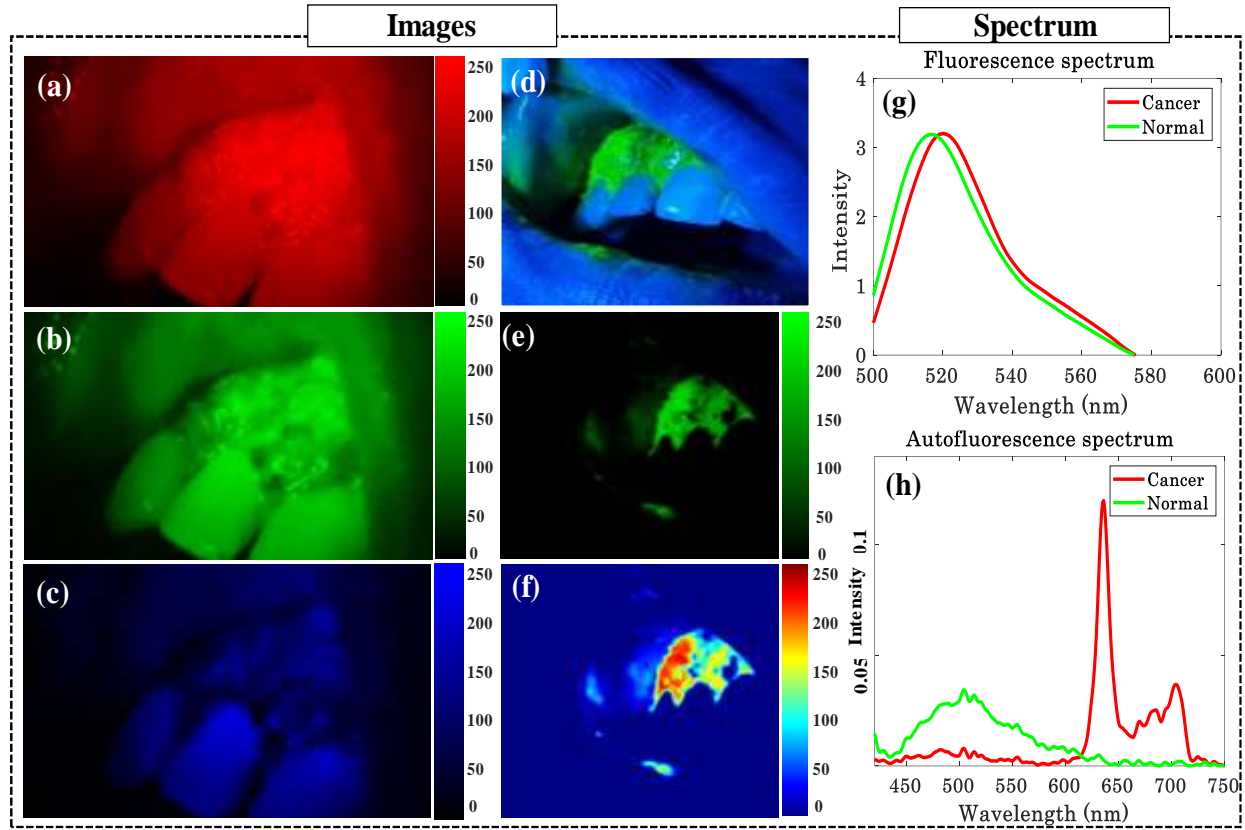

Figure 4. Autofluorescence and Fluorescence images and spectra of OSCC patient recorded at 365 nm (for AF images), 405 nm (for AF spectra) and 488 nm (for FL images and spectra). (a)-(c) Autofluorescence images of OSCC tissue in red, green, and blue region, respectively. (d)-(f) shows fluorescence images of same patient under the blue light, filtered image indicating tagging of dye in tissue region and intensity variation of tagged dye, respectively. (g) Autofluorescence spectra of same patient and (d) is fluorescence spectra of the patient.

Further, some pre-malignant patients of dysplasia and submucosa fibrosis are also visualized with present devices. Figure. 5 indicates AF and FL spectra of Dysplasia and submucosa fibrosis patients. Spectral features of dysplasia and fibrosis are distinguished from the OSCC. Figure 5(a) and (b) are showing AF and FL spectra of fibrosis, FL spectrum shows a shift of 1 nm and AF shows decrease of intensity in spectral-band compared to normal. Figure 5(c) and (d) are showing AF and FL spectra of dysplasia, FL spectrum shows a shift of 2 nm and AF shows decrease in spectral-band as well as a new peak formation around 554 nm. In fibrosis there is no such peaks in red region and any new peak formations as found in OSCC tissue and dysplastic tissue[32]. Generally, altered tissues (including cancerous and pre-cancerous tissues) has several degrees of blood content and keratinization in terms of tissue matrix molecules: NADH-FAD, porphyrins, etc., and thus could be reason for change in AF spectra[32, 33].

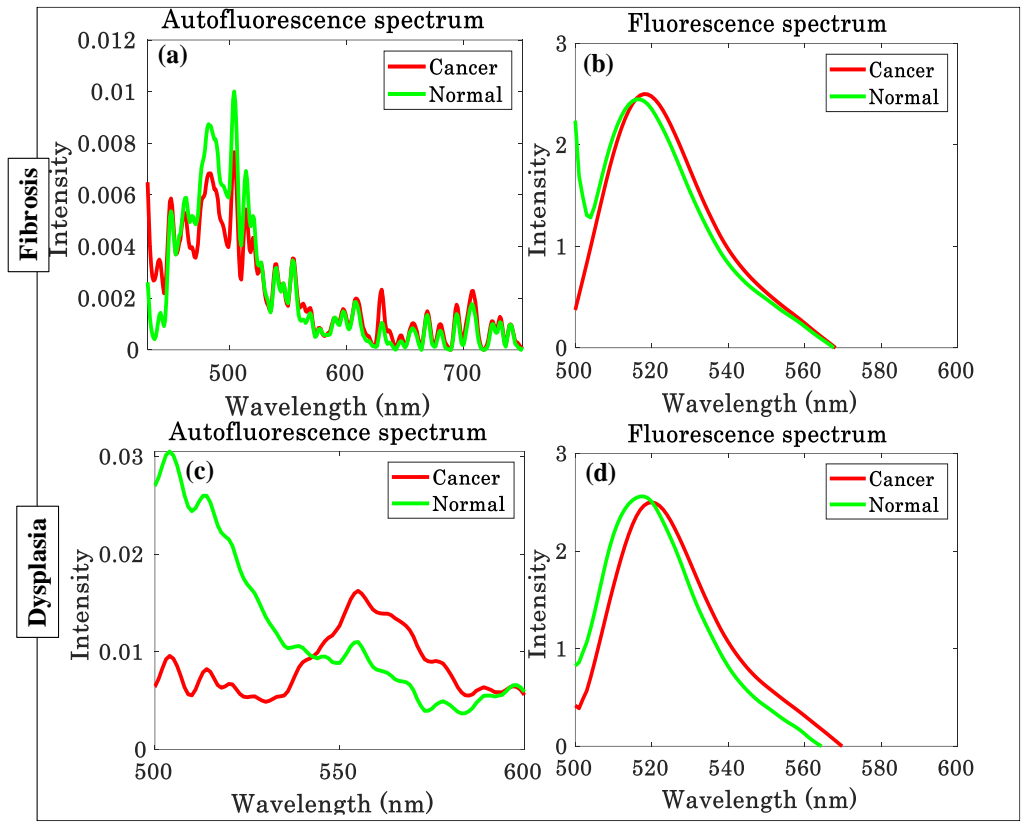

Figure 5. Autofluorescence and Fluorescence spectra for the dysplastic and fibrosis tissue. (a) Fluorescence spectrum of dysplasia, (b) is autofluorescence spectrum of same patient. (c) Fluorescence spectrum of fibrosis and (d) is autofluorescence spectrum of same patient.

Spectroscopy directly deals with molecular level and can be used as quantitative tool for identification of different cancerous stages. Observed spectra for OSCC and precancerous stages; dysplasia and fibrosis are shown in Fig.4 and Fig.5 and easily distinguishable in terms of red-shift, PpIX peaks and intensity ratio. At present, devices are used for classification of OSCC and normal but can be used for screening of precancerous stages also. A total 11 patients of mild hyperplasia with dysplasia and 2 patients of fibrosis are screened. For OSCC, average red shift is obtained about 6.59±4.54 nm, for dysplasia 3±4.78 nm and for fibrosis 1.5±0.5 nm. From observed data it is clearly visible that as we go towards higher grade of abnormality, emitted wavelength is more red-shifted due to change in its microenvironment[8, 34], detailed discussion is done in discussion section.

All features extracted from AF and FL images and spectra are used for classification of OSCC and normal tissue using SVM. A total 68 real sample data points from 18 patients are used for the classification, out of which 80% data is used for the training and 20% is used for testing. Figure 6 indicates ROC curve of testing data set, k-fold cross-validation is used for 5 times for training model and average sensitivity, specificity and accuracy are come to be 88.89%, 100 %, and 95%, respectively.

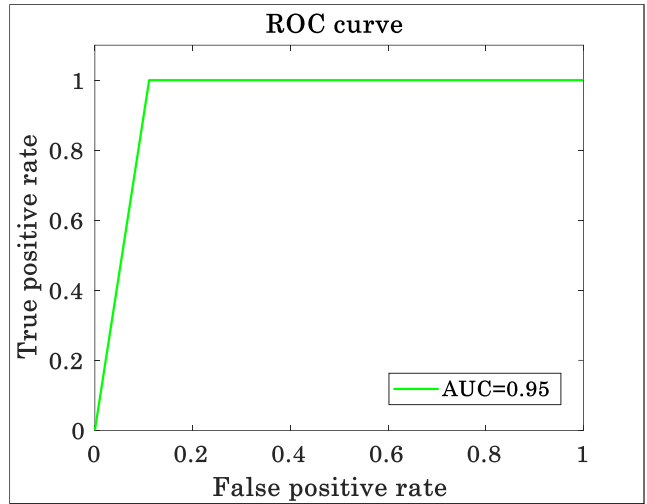

Figure 6. ROC curve for testing data set using SVM for OSCC tissue.

## 5. Discussion

FL and AF of cancerous tissue are different from normal tissues regarding biochemical and biophysical microenvironment of tissue architecture [16, 24, 33] and can be explained with Fig. 7 and following subsections. Figure. 7 (a) indicates mechanism of AF in oral tissue where hv is the incident energy, which changes due to different optical phenomena like absorption, scattering, reflection, energy transfer within tissue[34] giving AF (changed from incident energy) for different tissues; $hv_1$, $hv_2$, and $hv_3$ represents the emission in red, green, and blue channels respectively. Figure 7. (b) indicates oral tissue with FSS molecules. Neovascular surrounding of OSCC traps FSS molecules more against normal tissues. Here hv is incidence and $hv_1$ is emitted fluorescence from oral tissue.

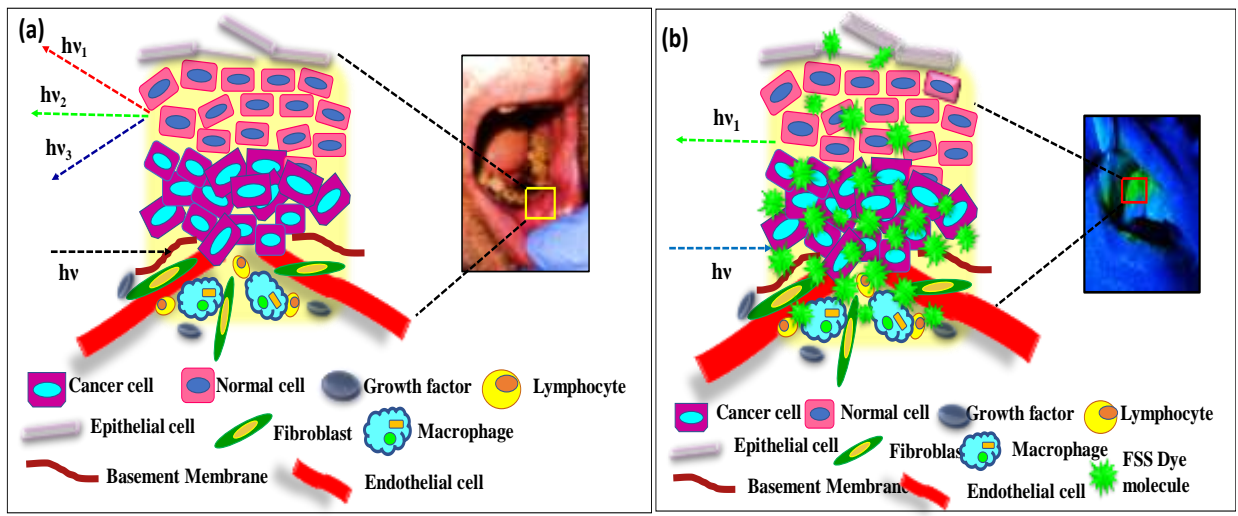

Figure 7. Schematic diagram of oral tissue. (a) Autofluorescence in oral tissue and black dotted line is showing the excitation wavelength to the tissue and blue, green, and red dotted lines are showing AF from the oral tissue. (b) Schematic of oral tissue after FSS dye application on it. Blue dotted line shows the excitation light to the tissue and green dotted line shows fluorescence emission.

### 5.1 Autofluorescence

AF of oral tissue shows different behavior for different conditions. NADH and FAD are complex compounds present in body and shows abatement of concentration in malignant tissues. Also, change in redox ratio in malignant tissues and healthy tissues indicates capability of AF in oral cancer screening. NADH and FAD are intracellular components and mainly participate in metabolic reactions of the cell and during amelioration of the disease, concentration of these complex compounds get changes and helps in identification of cancer evolution [18]. A spectral-band was observed around 430 nm- 560 nm for most of OSCC patients and AF was decreased compared to normal tissue. This spectral-band shows native fluorophores NADH, FAD and collagen crosslinks in oral tissue[18]. Progression of OSCC from normal tissue has significant variations which assigned to altered photo physical features of native fluorophores. Reduction of AF intensity in this spectral-band may be due to the disintegration of collagen crosslinks in stroma. As cancer tissues are heterogeneous which makes it more scattering media resulting in reduction of AF compared to normal tissues (can be seen in Fig S1. in supplementary file). Further NADH and FAD are primary electron donor and acceptor, respectively and plays role in metabolic reactions. Cancer cells restore their metabolism for proliferation, survival, growth and preservation within the tissue which results in shift from oxidative phosphorylation to glycolysis (energy generation through breakdown of glucose) in cancer cells[35]. Besides this, possible reasons for reduced AF are pH, solvation, oxidation state, and oxygenation concentration in tissue.

Apart from this spectral-band, significant differences between normal and OSCC tissue occurred in 630-710 nm spectral region. Three main peaks viz., ~636 nm, ~ 670 nm and ~ 706 nm observed in OSCC tissue which indicates free protoporphyrin IX (PpIX)[36]. Occurrence of these peaks in OSCC tissue is dependent on accumulation of porphyrins[36, 37]. Cancerous tissue indicates increased cellular metabolism, disintegration of RBC's (red blood cells) and increased neovascularization which helps in accumulation of porphyrins. Angiogenic developments (formation of new blood vessels) in cancerous tissues exhibit increased neovascularization which further increases blood content in superficial tissue layers. Cancerous tissue possesses hypervascularity which contains a lot of blood vessel than normal tissue resulting in more hemoglobin concentration in cancerous tissue. Hemoglobin's main component is porphyrin thus more hemoglobin indicates accumulation of porphyrin in cancerous tissue[37].

### 5.2 Fluorescence

Fluorescence shows high fluorescence intensity in cancerous tissue from normal and for cancerous tissue a significant red-shift is observed. Here dianion of FSS is used and FSS is pH dependent dye which gives more fluorescence in alkaline medium like aqueous humour[8]. Cancer cells intracellular pH value is alkaline in nature in comparison to normal cells resulting more fluorescence from cancerous tissues[38]. Also in cancerous tissue, there is formation of neovascular network which makes blood vessels very large in size with leaky capillaries due to this any dye applied to the tissue is captured into it and gives more fluorescence compared to normal tissue[38]. FSS is mainly make covalent bonds with two proteins, lysozyme and avidin; and lysozyme is mainly present in human saliva and notably present in oral cancer patient which leads more fluorescence in cancerous tissue[39, 40].

Another main and important difference is red-shift in cancerous tissue[8, 34]. Microenvironment of cancerous tissue shows more heterogeneity resulting varying refractive index which increases optical density in it shown in Fig.7. This optical density rises the optical path travelled by incident photon which has already interacted with FSS dye molecule thus giving a shifted emission[8, 34]. Also, this heterogeneity behaves as a highly scattering and absorbing media for incident photon which leads to increased random walk thus giving a red-shifted emission. Further, as discussed earlier AF also responsible for this red shift, as in cancerous tissue concentration of electron carriers (NADH or FAD) is changed[8].

## Conclusion

Present manuscript deals with screening of OSCC and precancerous oral stages with help of multimodal optical systems. Fluorescence and autofluorescence imaging and spectroscopy of oral tissue in real time is analyzed in present study. Fluorescence of oral tissue offers a significant red-shift for all stages while autofluorescence investigates native biomarkers in tissue. Adding both techniques within single study enhances the accuracy and real time reorganization/screening of oral cancer. A remarkable red-shift is recorded for all the stages and can be used as real time cancer marker, an average red-shift of 6.59±4.54 nm, 3±4.78 nm and 1.5±0.5 nm is obtained for OSCC, dysplasia and fibrosis, respectively. Further, SVM based classifier classifies the OSCC and normal tissues with the accuracy of 95% having sensitivity and specificity of 88.9 % and 100% respectively. Developed multimodal systems are field-portable, low cost and user friendly. Developing countries like India, which is hub for oral cancer; such devices are very beneficial as most of the patients are from rural India. At present, initial clinical trials of devices specify its capability for classifying the OSCC and identifying the precancerous stages. Further, devices can be used for a large scale of patients with different diagnosis and may give better results for different stages.

## Acknowledgment

The authors would like to acknowledge project MFIRP2019 No. IITD/IRD/M102229 for financial support. PT would like to acknowledge CSIR SRF fellowship with file No. 09/086(1338)/2018-EMR-I and Ankit Butola (Post Doc. at University of Tromso, Norway) for his valuable comments, assistantship and help in analysing the data set.

# References

[1] Markopoulos AK. Current aspects on oral squamous cell carcinoma. The open dentistry journal. 2012;6:126.

[2] Borse V, Konwar AN, Buragohain P. Oral cancer diagnosis and perspectives in India. Sensors International. 2020;1:100046.

[3] Thavarool SB, Muttath G, Nayanar S, Duraisamy K, Bhat P, Shringarpure K, et al. Improved survival among oral cancer patients: findings from a retrospective study at a tertiary care cancer centre in rural Kerala, India. World Journal of Surgical Oncology. 2019;17:1-7.

[4] Kumar P, Murali Krishna C. Optical techniques: investigations in oral cancers.  Oral Cancer Detection: Springer; 2019. p. 167-87.

[5] Huang T-T, Huang J-S, Wang Y-Y, Chen K-C, Wong T-Y, Chen Y-C, et al. Novel quantitative analysis of autofluorescence images for oral cancer screening. Oral oncology. 2017;68:20-6.

[6] Majumdar S, Uppal A, Gupta P. Autofluorescence spectroscopy of oral mucosa.  Optical Diagnostics of Biological Fluids III: International Society for Optics and Photonics; 1998. p. 158-68.

[7] Bhatt S, Butola A, Kanade SR, Kumar A, Mehta DS. High-resolution single-shot phase-shifting interference microscopy using deep neural network for quantitative phase imaging of biological samples. Journal of Biophotonics. 2021;14:e202000473.

[8] Thapa P, Singh V, Bhatt S, Tayal S, Mann P, Maurya K, et al. Development of multimodal micro-endoscopic system with oblique illumination for simultaneous fluorescence imaging and spectroscopy of oral cancer. Journal of Biophotonics. 2022:e202100284.

[9] Bailey MJ, Verma N, Fradkin L, Lam S, MacAulay CE, Poh CF, et al. Detection of precancerous lesions in the oral cavity using oblique polarized reflectance spectroscopy: a clinical feasibility study. Journal of Biomedical Optics. 2017;22:065002.

[10] Francisco ALN, Correr WR, Pinto CAL, Gonçalves Filho J, Chulam TC, Kurachi C, et al. Analysis of surgical margins in oral cancer using in situ fluorescence spectroscopy. Oral oncology. 2014;50:593-9.

[11] Francisco ALN, Correr WR, Azevedo LH, Kern VG, Pinto CAL, Kowalski LP, et al. Fluorescence spectroscopy for the detection of potentially malignant disorders and squamous cell carcinoma of the oral cavity. Photodiagnosis photodynamic therapy. 2014;11:82-90.

[12] Yaroslavsky AN, Feng X, Muzikansky A, Hamblin MR. Fluorescence polarization of methylene blue as a quantitative marker of breast cancer at the cellular level. Scientific reports. 2019;9:1-10.

[13] Dubey V, Ahmad A, Butola A, Qaiser D, Srivastava A, Mehta DS. Low coherence quantitative phase microscopy with machine learning model and Raman spectroscopy for the study of breast cancer cells and their classification. Applied optics. 2019;58:A112-A9.

[14] Butola A, Ahmad A, Dubey V, Srivastava V, Qaiser D, Srivastava A, et al. Volumetric analysis of breast cancer tissues using machine learning and swept-source optical coherence tomography. Applied optics. 2019;58:A135-A41.

[15] Tayal S, Singh V, Kaur T, Singh N, Mehta DS. Simultaneous fluorescence and quantitative phase imaging of MG63 osteosarcoma cells to monitor morphological changes with time using partially spatially coherent light source. Methods Applications in Fluorescence. 2020;8:035004.

[16] Kumar P, Kanaujia SK, Singh A, Asima P. In vivo detection of oral precancer using a fluorescence-based, in-house-fabricated device: a Mahalanobis distance-based classification. Lasers in medical science. 2019;34:1243-51.

[17] Poh CF, MacAulay CE, Zhang L, Rosin MP. Tracing the "at-risk" oral mucosa field with autofluorescence: steps toward clinical impact. Cancer Prevention Research. 2009;2:401-4.

[18] Skala MC, Riching KM, Gendron-Fitzpatrick A, Eickhoff J, Eliceiri KW, White JG, et al. In vivo multiphoton microscopy of NADH and FAD redox states, fluorescence lifetimes, and cellular morphology in precancerous epithelia. Proceedings of the National Academy of Sciences. 2007;104:19494-9.

[19] VELscope, Enhanced Oral Assesment.

[20] OralScan, Oral Cancer Screening Device.

[21] Awan K, Morgan P, Warnakulasuriya S. Evaluation of an autofluorescence based imaging system (VELscope™) in the detection of oral potentially malignant disorders and benign keratoses. Oral oncology. 2011;47:274-7.

[22] Scheer M, Fuss J, Derman MA, Kreppel M, Neugebauer J, Rothamel D, et al. Autofluorescence imaging in recurrent oral squamous cell carcinoma. Oral maxillofacial surgery. 2016;20:27-33.

[23] Kim DH, Kim SW, Hwang SH. Autofluorescence imaging to identify oral malignant or premalignant lesions: Systematic review and meta-analysis. Head Neck. 2020;42:3735-43.

[24] Ramanujam N. Fluorescence spectroscopy in vivo. Encyclopedia of analytical chemistry. 2000;1:20-56.

[25] Riaz A, Shreedhar B, Kamboj M, Natarajan S. Methylene blue as an early diagnostic marker for oral precancer and cancer. Springerplus. 2013;2:1-7.

[26] Lakowicz JR. Principles of fluorescence spectroscopy: Springer; 2006.
[27] Huang T-T, Chen K-C, Wong T-Y, Chen C-Y, Chen W-C, Chen Y-C, et al. Two-channel autofluorescence analysis for oral cancer. Journal of biomedical optics. 2018;24:051402.

[28] Yuvaraj M, Udayakumar K, Jayanth V, Rao AP, Bharanidharan G, Koteeswaran D, et al. Fluorescence spectroscopic characterization of salivary metabolites of oral cancer patients. Journal of Photochemistry Photobiology B: Biology. 2014;130:153-60.

[29] OLSON JL, MANDAVA N. Fluorescein angiography. Retinal Imaging: Elsevier; 2006. p. 3-21.

[30] Qaiser D, Sood A, Mishra D, Kharbanda O, Srivastava A, Gupta SD, et al. Novel use of fluorescein dye in detection of oral dysplasia and oral cancer. Photodiagnosis Photodynamic Therapy. 2020;31:101824.

[31] Huang S, Cai N, Pacheco PP, Narrandes S, Wang Y, Xu W. Applications of support vector machine (SVM) learning in cancer genomics. Cancer genomics proteomics. 2018;15:41-51.

[32] de Veld DC, Skurichina M, Witjes MJ, Duin RP, Sterenborg HJ, Roodenburg JL. Clinical study for classification of benign, dysplastic, and malignant oral lesions using autofluorescence spectroscopy. Journal of biomedical optics. 2004;9:940-50.

[33] Haris P, Balan A, Jayasree R, Gupta A. Autofluorescence spectroscopy for the in vivo evaluation of oral submucous fibrosis. Photomedicine Laser Surgery. 2009;27:757-61.

[34] Ghasemi F, Parvin P, Motlagh NSH, Abachi S. LIF spectroscopy of stained malignant breast tissues. Biomedical Optics Express. 2017;8:512-23.

[35] Liberti MV, Locasale JW. The Warburg effect: how does it benefit cancer cells? Trends in biochemical sciences. 2016;41:211-8.

[36] Gurushankar K, Nazeer SS, Gohulkumar M, Jayasree RS, Krishnakumar N. Endogenous porphyrin fluorescence as a biomarker for monitoring the anti-angiogenic effect in antitumor response to hesperetin loaded nanoparticles in experimental oral carcinogenesis. RSC Advances. 2014;4:46896-906.

[37] Palmer S, Litvinova K, Dunaev A, Yubo J, McGloin D, Ghulam N. Optical redox ratio and endogenous porphyrins in the detection of urinary bladder cancer: A patient biopsy analysis. Journal of Biophotonics. 2017;10:1062-73.

[38] Hao G, Xu ZP, Li L. Manipulating extracellular tumour pH: an effective target for cancer therapy. RSC advances. 2018;8:22182-92.

[39] Jityuti B, Kuno M, Liwporncharoenvong T, Buranaprapuk A. Selective protein photocleavage by fluorescein derivatives. Journal of Photochemistry Photobiology B: Biology. 2020;212:112027.

[40] Sun H, Chen Y, Zou X, Li Q, Li H, Shu Y, et al. Salivary secretory immunoglobulin (SIgA) and lysozyme in malignant tumor patients. BioMed Research International. 2016;2016.